\documentclass[twocolumn,showpacs,showkeys]{revtex4}
\usepackage{graphicx}
\usepackage{bm}
\usepackage{color}
\usepackage{amsmath}
\usepackage{natbib}

\begin{document}

\title{Extraordinary waves in two dimensional electron gas with separate spin evolution and Coulomb exchange interaction}

\author{Pavel A. Andreev}
\email{andreevpa@physics.msu.ru} \affiliation{Faculty of physics,
Lomonosov Moscow State University, Moscow, Russian Federation.}

 \date{\today}

\begin{abstract}

Hydrodynamics analysis of waves in two-dimensional degenerate
electron gas with the account of separate spin evolution is
presented. The transverse electric field is included along with
the longitudinal electric field. The Coulomb exchange interaction
is included in the analysis. In contrast with the
three-dimensional plasma-like mediums the contribution of the
transverse electric field is small. We show the decrease of
frequency of both the extraordinary (Langmuir) wave and the
spin-electron acoustic wave due to the exchange interaction.
Moreover, spin-electron acoustic wave has negative dispersion at
the relatively large spin-polarization. Corresponding dispersion
dependencies are presented and analyzed.
\end{abstract}


\maketitle








\section{\label{sec:level1} Introduction}

Collective excitations in spin-polarized electron gas are
considered as waves propagating in a plasma-like medium. The
quantum hydrodynamic (QHD) method \cite{MaksimovTMP 1999},
\cite{MaksimovTMP 2001} and its generalization for the separate
spin evolution (SSE) \cite{Andreev PRE 15 SEAW} are applied for
the theoretical modeling of this system.

The SSE-QHD method is developed for the spin-polarized degenerate
electron gas \cite{Andreev PRE 15 SEAW}. Majority of papers on
this subject are focused on three-dimensional electron gas
existing in magnetically ordered metals (see for instance
\cite{Andreev AoP 15 SEAW}, \cite{Andreev 1512 surface}). However,
it can be applied to the spin-polarized electron-hole liquid in
the regime of degenerate carriers. It can be done similarly to the
analysis of the electron-positron plasmas \cite{Andreev PRE 16}.
Including difference of the effective masses of the electrons and
holes or (and) different concentrations of the electrons and
holes, we find more similarity between electron-hole liquid and
electron-positron-ion plasmas.  Approximately, the SSE-QHD may be
applied for non-degenerate electron-hole liquid, but it requires a
modified equation of state and account of the scattering processes
leading to the damping of excitations.

The SSE-QHD allows to consider linear \cite{Andreev AoP 15 SEAW},
\cite{Andreev PRE 16} and nonlinear \cite{Andreev PoP 16 exchange}
bulk spin-electron acoustic waves (SEAWs). It also allows to
consider the surface SEAWs \cite{Andreev 1512 surface} and SEAWs
in two-dimensional electron gas \cite{Andreev EPL 16}.

Spectrum of the plasma-like mediums is highly affected by the
transverse electric field at the wave propagation perpendicular to
the external field. The transverse electric field leads to
formation of the extraordinary waves instead of the Langmuir or
upper hybrid waves for the electrostatic regime \cite{Landau v10}.
Same effect was demonstrated for the SEAWs \cite{Trukhanova 1603}.
Therefore, we pay special attention to the transverse electric
field in two-dimensional electron gas.

The collective excitations in the two-dimensional electron gas
with no account of the spin polarization have been studied for a
long time \cite{Stern PRL 67}, \cite{Grimes PRL 76}, including the
magnetoplasmons existing in the two-dimensional electron gas
located in the external magnetic field \cite{Oji PRB 86},
\cite{Batke PRB 86}. The spin-polarized two-dimensional electron
gas exists in the magnetic semiconductors. The study of SEAWs in
two-dimensional electron gas appeared by analogy with bulk
properties of the three-dimensional electron gas \cite{Andreev PRE
15 SEAW}, \cite{Andreev AoP 15 SEAW}. However, there is a
relatively long history of theoretical and experimental study of
the spin plasmons in he spin-polarized two-dimensional electron
gas (see for instance \cite{Ryan PRB 91}, \cite{Voitenko JLTP
95}). The spin plasmons show dispersion and physical mechanism to
the SEAWs. This similarity shows that described above linear and
nonlinear, bulk and surface SEAWs can be considered as spin
plasmons in three-dimensional electron gas.

It seems that all analysis of spin plasmons \cite{Ryan PRB 91} is
focused on the quantum wells in the magnetized semiconductors.
Acoustic plasmon frequency is experimentally measured and the
results are compared with the values calculated for spin polarized
electrons and holes in p-type $A_{3}B_{5}$ semiconductors
\cite{Voitenko JLTP 95}. Spin flip waves and spin density
fluctuations of a two-dimensional spin-polarized electron-gas in a
semimagnetic $Cd_{1–0.008}Mn_{0.008}Te$ quantum well are
considered in Ref. \cite{Perez PSS 06}. A maximum value of the
spin polarization degree of 35 percent is deduced from
measurements. Collective excitations in the spin polarized quantum
well were also considered in Refs. \cite{Marinescu PRB 98},
\cite{Yi Phys E 00}, \cite{Perez PRB 09}, \cite{Barate PRB 10}.
Recently, the contribution of Rashba spin-orbit coupling in the
collective modes in two- and three-dimensional electron systems
considered in Refs. \cite{Ullrich PRB 02}, \cite{Maiti PRB 15}.

Large contribution of the exchange interaction in the
two-dimensional electron gas of GaAs microstructures is
demonstrated experimentally \cite{Pinczuk PRL 89} and
theoretically within the local density approximation \cite{Bloss
JAP 89}. Hence, we consider the Coulomb exchange interaction
contribution in spectrum of the Langmuir waves (plasmons) and the
SEAWs (spin plasmons) in terms of the SSE-QHD. The influence of
the exchange interaction on the collective effects in a quantum
well was analyzed in Ref. \cite{Ryan PRB 91 b}.

\section{\label{sec:level1} Model}

We apply the SSE-QHD model developed in Ref. \cite{Andreev PRE 15
SEAW}, \cite{Andreev AoP 15 SEAW} and adopted for two-dimensional
systems in Ref. \cite{Andreev EPL 16}. Since we are interested in
analysis of the effects of Coulomb exchange interaction we use
generalization of the SSE-QHD containing the exchange interaction
\cite{Andreev PoP 16 exchange}. However, Ref. \cite{Andreev PoP 16
exchange} contains the exchange interaction for the
three-dimensional electron gas. We need to substitute it in the
similar way by the exchange interaction in two-dimensional
electron gas in accordance with Refs. \cite{Datta JAP 83},
\cite{Andreev AoP 14 exchange}. We also use the full integral
Maxwell equations for the expressions of the electric and magnetic
fields to consider the contribution of the transverse electric
field in the waves in magnetized spin-polarized electron gas.
Overall we have the following set of four hydrodynamic equations:
the continuity equations
\begin{equation}\label{2DExWsse cont eq electrons spin UP}
\partial_{t}n_{s}+\nabla(n_{s}\textbf{v}_{s})=\pm\frac{\gamma}{\hbar}(S_{x}B_{y}-S_{y}B_{x}), \end{equation}
and the Euler equations
$$mn_{s}(\partial_{t}+\textbf{v}_{s}\nabla)\textbf{v}_{s}+\nabla p_{s}$$
\begin{equation}\label{2DExWsse Euler eq spin UP with vel} =q_{e}n_{s}\biggl(\textbf{E}+\frac{1}{c}[\textbf{v}_{s},\textbf{B}]\biggr)+\textbf{F}_{SS,s},\end{equation}
with the force field of spin-spin interaction
$$\textbf{F}_{SS,s}= \pm\gamma_{e}n_{s}\nabla B_{z} +\frac{\gamma_{e}}{2}(S_{x}\nabla B_{x}+S_{y}\nabla B_{y})$$
\begin{equation}\label{2DExWsse} \pm\frac{m\gamma_{e}}{\hbar}[(\textbf{J}_{(M)x}-\textbf{v}_{s}S_{x})B_{y}-(\textbf{J}_{(M)y}-\textbf{v}_{s}S_{y})B_{x}],\end{equation}
containing the spin current
\begin{equation}\label{2DExWsse Spin current x} \textbf{J}_{(M)\alpha}=\frac{1}{2}(\textbf{v}_{u}+\textbf{v}_{d})S_{\alpha}-\varepsilon_{\alpha\beta z}\frac{\hbar}{4m} \biggl(\frac{\nabla n_{u}}{n_{u}}-\frac{\nabla n_{d}}{n_{d}}\biggr)S_{\beta}, \end{equation}
and the effective pressure
\begin{equation}\label{2DExWsse Eq State partial} p_{s}=\pi\hbar^{2}n_{s}^{2}/m-\zeta  \frac{8\beta}{3\pi \sqrt{\pi}}q_{e}^{2}n_{d}^{\frac{3}{2}}\delta_{sd},\end{equation}
where we introduce $\beta\equiv 24
\textrm{arsh}1=24\ln(1+\sqrt{2})=21.153$ and
\begin{equation}\label{EXCHANGE} \zeta=1-\frac{(1-\eta)^{3/2}}{(1+\eta)^{3/2}}.\end{equation}

The electromagnetic field is caused by the charges, electric
currents and magnetic moments (spin) $\textbf{E}=-\nabla
\varphi-\partial_{t}\textbf{A}/c$,
$\textbf{B}=\textbf{B}_{ext}+\nabla\times
\textbf{A}_{\textbf{j}}+\textbf{B}_{spin}$, where
\begin{equation}\label{2DExWsse } \varphi(\textbf{r},t)= \sum_{s=u,d} q_{e}\int\frac{n_{s}(\textbf{r}',t-|\textbf{r}-\textbf{r}'|/c)}{|\textbf{r}-\textbf{r}'|}d\textbf{r}',\end{equation}
\begin{equation}\label{2DExWsse } \textbf{A}_{\textbf{j}}(\textbf{r},t)= \sum_{s=u,d} \frac{q_{e}}{c}\int\frac{\textbf{j}_{s}(\textbf{r}',t-|\textbf{r}-\textbf{r}'|/c)}{|\textbf{r}-\textbf{r}'|}d\textbf{r}',\end{equation}
and
\begin{equation}\label{2DExWsse} \textbf{B}_{spin}(\textbf{r},t)=\int  [(\textbf{M}(\textbf{r}',t')\nabla)\nabla-\textbf{M}(\textbf{r}',t')\triangle]\frac{1}{\mid \textbf{r}-\textbf{r}'\mid}d\textbf{r}',\end{equation}
where
$\textbf{M}(\textbf{r}',t')=\textbf{M}(\textbf{r}',t-|\textbf{r}-\textbf{r}'|/c)$
is the magnetization existed in point with coordinate
$\textbf{r}'$ in an earlier moment of time $t'$, with
$d\textbf{r}'=dx'dy'$ is the element of volume in 2D space.

\section{\label{sec:level1} Linearized equations}

We consider the propagation of plane waves along the Ox direction:
$\textbf{k}=\{ k_{x}, 0, 0\}$. In the longitudinal waves (the
electrostatic waves), the perturbation of electric field is
parallel to the direction of wave propagation. However, we
consider the longitudinally-transverse waves. Hence, we include
the electric field perturbation along the Oy direction:
$\textbf{E}=\{ E_{x}, E_{y}, 0\}$.

\begin{equation}\label{2DExWsse cont lin 2 fl} -\imath\omega\delta n_{s}+\imath k_{x} n_{0s}\delta v_{sx}=0,\end{equation}
\begin{equation}\label{2DExWsse Euler lin 2 fl} -\imath\omega mn_{0s}\delta \textbf{v}_{s}+\imath \textbf{k}\delta p_{s}=q_{e}n_{0s}\delta \textbf{E}+mn_{0s}\Omega_{e}[\delta \textbf{v}_{s}, \textbf{e}_{z}],\end{equation}
where $\Omega_{e}=q_{e}B_{0}/mc$ is the cyclotron frequency. We do
not include the spin-spin interaction force field in the Euler
equation as a small effect.

The perturbation of electric field has the following connection
with the perturbation of concentration and the velocity field
$$\delta \textbf{E}=-q_{e}\sum_{s=u,d}\biggl(\nabla\int\frac{\delta n_{s}(\textbf{r}',t-|\textbf{r}-\textbf{r}'|/c)}{|\textbf{r}-\textbf{r}'|}d\textbf{r}' $$
\begin{equation}\label{2DExWsse } +\frac{n_{0s}}{c^{2}}\partial_{t}\int\frac{\delta\textbf{v}_{s}(\textbf{r}',t-|\textbf{r}-\textbf{r}'|/c)}{|\textbf{r}-\textbf{r}'|}d\textbf{r}' \biggr).\end{equation}

As it is demonstrated in Appendix for plane waves, the
perturbation of the electric field can be presented in the
following form:
\begin{equation}\label{2DExWsse } \delta \textbf{E}=\imath q_{e}\sum_{s=u,d}\biggl(-\Im \textbf{k}\delta n_{s}
+\Im \frac{\omega}{c^{2}}n_{0s}\delta
\textbf{v}_{s}\biggr).\end{equation}

\begin{figure}
\includegraphics[width=8cm,angle=0]{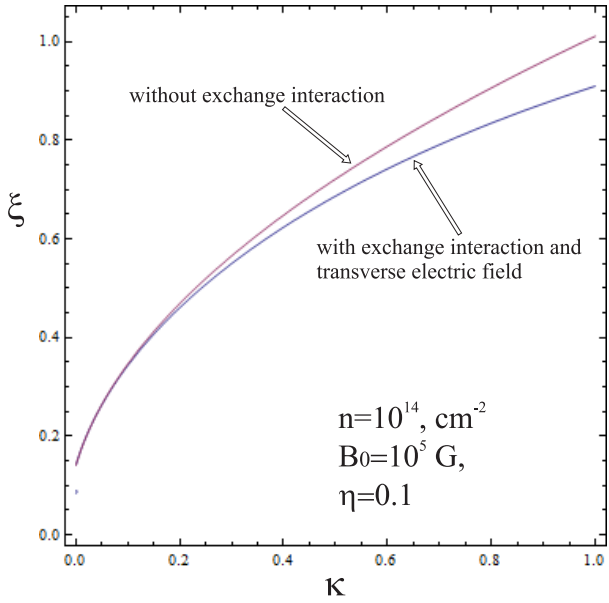}
\caption{\label{2DExWsse_SF_01} (Color online) The figure shows
the dimensionless dispersion dependence $\omega/\omega_{0}$ on the
dimensionless wave vector $\kappa=k/\sqrt{n_{0}}$, where
$\omega_{0}^{2}=2\pi e^{2}n_{0e}^{3/2}/m$ is a constant giving a
characteristic frequency. Parameters are presented in the figure.}
\end{figure}
\begin{figure}
\includegraphics[width=8cm,angle=0]{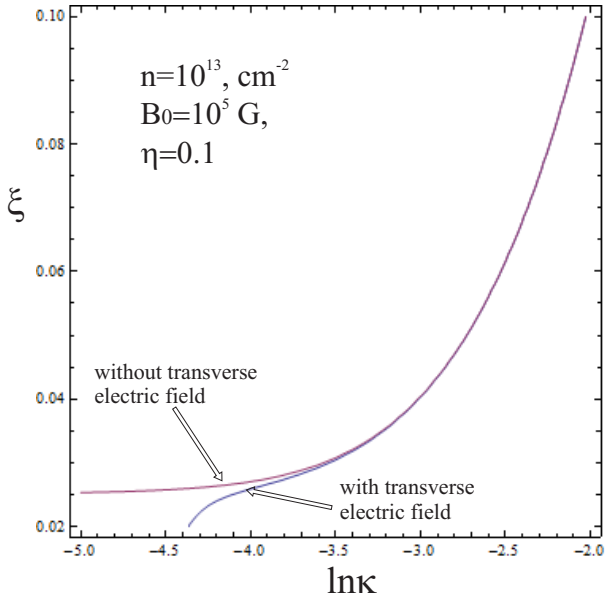}
\caption{\label{2DExWsse_SF_01} (Color online) The figure shows a
considerable change of the dispersion dependence at the small wave
vectors due to the transverse electric field of wave. We present
the dimensionless frequency $\xi$ as a function of the natural
logarithm of the dimensionless wave vector $\ln\kappa$.}
\end{figure}

As we will see below the result of application of equations
(\ref{2DExWsse cont lin 2 fl}) and (\ref{2DExWsse Euler lin 2 fl})
is rather complicate. Thus, as an intermediate step we consider
the extraordinary waves in 2DEG in terms of the single fluid model
of electrons.

\section{Single fluid model of electrons and extraordinary wave dispersion}

Linearized set of quantum hydrodynamic equations for electrons
considered as a single fluid is:
\begin{equation}\label{2DExWsse SF cont linear} -\imath\omega\delta n_{e}+\imath k_{x} n_{0e}\delta v_{ex}=0, \end{equation}
\begin{equation}\label{2DExWsse } -\imath\omega n_{0e}\delta v_{ex}+\imath k_{x}U_{e}^{2}\delta n_{e}=\frac{q_{e}}{m}n_{0e}\delta E_{x}+n_{0e}\Omega_{e}\delta v_{ey}, \end{equation}
\begin{equation}\label{2DExWsse } -\imath\omega n_{0e}\delta v_{ey}=\frac{q_{e}}{m}n_{0e}\delta E_{y}-n_{0e}\Omega_{e}v_{ex},\end{equation}
where
\begin{equation}\label{2DExWsse } \delta E_{x}=q_{e}\Im\imath\biggl(-k_{x}\delta n_{e}+\frac{\omega}{c^{2}}n_{0e}\delta v_{x}\biggr),\end{equation}
\begin{equation}\label{2DExWsse SF eq for Ey} \delta E_{y}=q_{e}\Im\imath \frac{\omega}{c^{2}}n_{0e}\delta v_{y},\end{equation}
and
$U_{e}^{2}=(1+\eta^{2})\frac{\pi\hbar^{2}n_{0e}}{m_{e}^{2}}-(1+\eta)^{3/2}\zeta\frac{\beta\sqrt{2\pi}e^{2}}{\pi^{2}m_{e}}\sqrt{n_{0e}}$.

Equations (\ref{2DExWsse SF cont linear})-(\ref{2DExWsse SF eq for
Ey}) lead to the following dispersion equation for the
longitudinally-transverse waves in spin-polarized two-dimensional
electron gas
$$\omega^{2}\biggl(1+\frac{1}{\sqrt{1-\frac{\omega^{2}}{k^{2}c^{2}}}}\frac{\omega_{Le}^{2}}{k^{2}c^{2}}\biggr)-k^{2}U_{e}^{2}$$
\begin{equation}\label{2DExWsse disp eq SF with TrF} -\frac{1}{\sqrt{1-\frac{\omega^{2}}{k^{2}c^{2}}}}\omega_{Le}^{2} =\frac{\Omega_{e}^{2}}{1+\frac{1}{\sqrt{1-\frac{\omega^{2}}{k^{2}c^{2}}}}\frac{\omega_{Le}^{2}}{k^{2}c^{2}}},\end{equation}
where we have used the two dimensional Langmuir frequency
\begin{equation}\label{2DExWsse Langmuir frq 2D}\omega_{Le}^{2}=\frac{2\pi e^2 k n_{0e}}{m}\sim k.\end{equation}

To drop the transverse electric field contribution we should
consider the limit case $c\rightarrow\infty$. Thus, we find the
dispersion dependence for the longitudinal waves in 2DEG in the
single fluid model of electrons (see for instance \cite{Andreev
AoP 14 exchange})
\begin{equation}\label{2DExWsse disp Lang 2D with ex} \omega^{2}=\Omega_{e}^{2}+\omega_{Le}^{2}+U_{e}^{2}k^{2}. \end{equation}

In both regimes we find single branch of the dispersion
dependence.

Considering the linear perturbations we can compare the
contributions of the Fermi pressure and the exchange interaction.
Their relative behavior can be described by the following
dimensionless parameter:
$\Lambda=\frac{\hbar^{2}}{me^{2}}\sqrt{n_{0e}}$ \cite{Andreev AoP
14 exchange}. More precisely, the Fermi pressure is larger then
the Coulomb exchange interaction if the concentrations of
electrons satisfy the following condition
$\sqrt{n_{0d}}>\frac{2\zeta\sqrt{\pi}\beta}{\pi^{3}(1+\eta^{2})}\frac{me^{2}}{\hbar^{2}}$.
It gives $n_{0d}(\eta=0.1)=6\times10^{16}$ cm$^{-2}$ or
$n_{0d}(\eta=0.01)=7\times10^{15}$ cm$^{-2}$.

\begin{figure}
\includegraphics[width=8cm,angle=0]{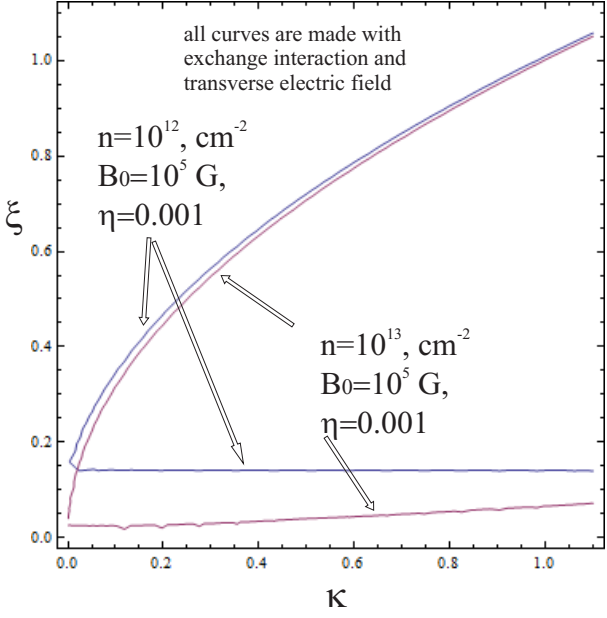}
\caption{\label{2DExWsse_SF_01} (Color online) The figure shows
solutions of dispersion equation (\ref{2DExWsse Disp Eq two Fluids
with Tr field}) for regimes of two different concentrations at the
fixed spin polarization. In each regime we find two solutions: the
Langmuir wave and the SEAW.}
\end{figure}
\begin{figure}
\includegraphics[width=8cm,angle=0]{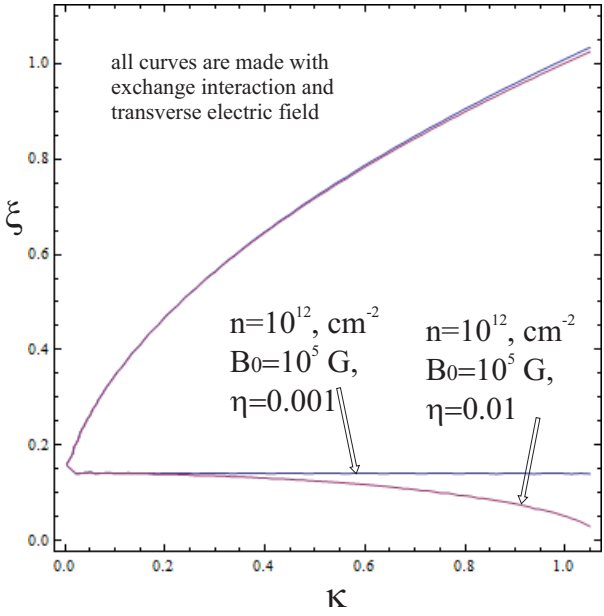}
\caption{\label{2DExWsse_SF_01} (Color online) The figure shows
dispersion dependencies arising as solutions of equation
(\ref{2DExWsse Disp Eq two Fluids with Tr field}) for regimes of
two different spin polarizations at the same concentrations and
external magnetic field. We assume that inner interaction with
ions of the medium give dominant contribution to the spin
polarization.}
\end{figure}

Numerical analysis shows that the change of the dispersion
dependence due to the transverse electric field is very small, it
arises in the regime of small wave vectors only. Formulae
(\ref{2DExWsse disp Lang 2D with ex}) and (\ref{2DExWsse disp eq
SF with TrF}) gives coinciding curves in Fig. 1. The small change
of spectrum exists at the small wave vectors, as it is
demonstrated in Fig. 2. Comparing two curves in Fig. 1 we see that
the Coulomb exchange interaction gives considerable decrease of
the dispersion dependence of the Langmuir waves or the
extraordinary waves with the small contribution of the transverse
electric field.

\section{Dispersion dependence for two-fluid model of electrons}

After the calculation of electromagnetic potentials the linear SSE
hydrodynamic equations (\ref{2DExWsse cont lin 2 fl}),
(\ref{2DExWsse Euler lin 2 fl}) can be presented in a local form
as a set of algebraic equations:
\begin{equation}\label{2DExWsse concentr perturb 2 fluids} \delta n_{s}=n_{0s}k_{x}\delta v_{xs}/\omega,\end{equation}
$$-\imath\omega mn_{0s}\delta v_{sx}+\imath k_{x} mU_{s}^{2}\delta n_{s}=q_{e}^{2}n_{0s}\imath\Im \biggl(-k_{x}\delta n_{u}-k_{x}\delta n_{d}$$
\begin{equation}\label{2DExWsse v x perturb 2 fluids} +\frac{\omega}{c^{2}}n_{0u}\delta v_{ux}+\frac{\omega}{c^{2}}n_{0d}\delta v_{dx}\biggr)+mn_{0s}\Omega_{e}\delta v_{sy},\end{equation}
$$-\imath\omega mn_{0s}\delta v_{sy}=q_{e}^{2}n_{0s}\imath\Im \biggl(\frac{\omega}{c^{2}}n_{0u}\delta v_{uy}$$
\begin{equation}\label{2DExWsse v y perturb 2 fluids} +\frac{\omega}{c^{2}}n_{0d}\delta v_{dy}\biggr) -mn_{0s}\Omega_{e}\delta v_{sx},\end{equation}
where
$U_{s}^{2}=\frac{2\pi\hbar^{2}}{m^{2}}n_{0s}-\zeta\frac{4\beta
e^{2}}{\pi\sqrt{\pi}m_{e}}\sqrt{n_{0d}}\delta_{sd}$.

Substituting formula (\ref{2DExWsse concentr perturb 2 fluids}) to
equations (\ref{2DExWsse v x perturb 2 fluids}) and (\ref{2DExWsse
v y perturb 2 fluids}) we obtain a set of four uniform algebraic
equations. This set has a nonzero solution if its determinant is
equal to zero. It leads to the following dispersion equation:
\begin{widetext}
$$\biggl(1+\frac{\omega_{Ru}^{2}+\omega_{Rd}^{2}}{k^{2}c^{2}}\biggr)\biggl[(\omega^{2}-k^{2}U_{u}^{2})(\omega^{2}-k^{2}U_{d}^{2})
-[\omega_{Ru}^{2}(\omega^{2}-k^{2}U_{d}^{2})+\omega_{Rd}^{2}(\omega^{2}-k^{2}U_{u}^{2})]\biggl(1-\frac{\omega^{2}}{k^{2}c^{2}}\biggr)\biggr]$$
\begin{equation}\label{2DExWsse Disp Eq two Fluids with Tr field} -\Omega_{e}^{2}\Biggl\{\biggl[\omega^{2}-k^{2}U_{u}^{2}-\omega_{Ru}^{2}\biggl(1-\frac{\omega^{2}}{k^{2}c^{2}}\biggr)\biggr]\biggl(1+\frac{\omega_{Ru}^{2}}{k^{2}c^{2}}\biggr)
+\biggl[\omega^{2}-k^{2}U_{d}^{2}-\omega_{Rd}^{2}\biggl(1-\frac{\omega^{2}}{k^{2}c^{2}}\biggr)\biggr]\biggl(1+\frac{\omega_{Rd}^{2}}{k^{2}c^{2}}\biggr)\Biggr\}+\Omega_{e}^{4}=0,\end{equation}
\end{widetext}
where
$\omega_{Rs}^{2}=q_{e}^{2}n_{0s}k^{2}\Im/m=\omega_{Ls}^{2}/\sqrt{1-\omega^{2}/k^{2}c^{2}}$
is the retarding Langmuir frequency.

In the electrostatic limit ($c\rightarrow\infty$) we can drop the
transverse electric field contribution in equation (\ref{2DExWsse
Disp Eq two Fluids with Tr field}) and find the following
dispersion equation:
\begin{equation}\label{2DExWsse Disp Eq two Fluids longit} (\omega^{2}-\omega_{Lu}^{2}-\Omega_{e}^{2}-k^{2}U_{u}^{2})
(\omega^{2}-\omega_{Ld}^{2}-\Omega_{e}^{2}-k^{2}U_{d}^{2})
-\omega_{Lu}^{2}\omega_{Ld}^{2}=0.\end{equation} It was derived in
Ref. \cite{Andreev EPL 16} (the dispersion equation is not
presented in the paper, but its solution is presented by formula
7), it is also similar to results of earlier Ref. \cite{Ryan PRB
91}, starting from the SSE-QHD in the electrostatic regime.

Both equations (\ref{2DExWsse Disp Eq two Fluids with Tr field})
and (\ref{2DExWsse Disp Eq two Fluids longit}) have two solutions
describing two branches of the wave dispersion dependence: the
Langmuir (plasmon) mode and the spin-electron acoustic mode
(spin-plasmon mode). Equation (\ref{2DExWsse Disp Eq two Fluids
with Tr field}) describes these waves in the regime of
longitudinally-transverse waves. In this case, similarly to
three-dimensional plasma-like mediums, we can call these wave the
extraordinary waves.

At the account of the SSE we find numerically negligibly small
contribution of the transverse electric field. Hence, equations
(\ref{2DExWsse Disp Eq two Fluids with Tr field}) and
(\ref{2DExWsse Disp Eq two Fluids longit}) give coinciding curves
in Figs. 3, 4.

In Fig. 3 we see that at larger concentrations the dimensionless
frequency of the Langmuir waves (the pair of upper curves) has
smaller value and growth rate. The SEAWs (the pair of lower
curves) have considerable smaller dimensionless frequencies at the
larger concentration, but the growth rate becomes larger at the
larger concentration.

As we see from Fig. 4 the increase of the spin polarization leads
to the decrease of the frequencies of both waves. This effect
reveals itself more in the SEAW spectrum. The exchange interaction
decreases the SEAW frequency down to negative group velocity.
Further increase of the spin polarization decreases the area of
SEAW existence since its frequency goes down to zero value at the
accessible wave vectors $\kappa <1$.

\section{\label{sec:level1} Conclusion}

We have considered waves in the magnetized two-dimensional
electron gas located in the external constant uniform magnetic
field directed perpendicular to the plane. We have payed attention
to both the longitudinal and the transverse parts of the electric
field of the wave perturbation propagating in the spin-polarized
degenerate two-dimensional electron gas. Considering all electrons
as the single fluid we find one wave solution: the extraordinary
wave or the hybrid wave in the electrostatic limit. It was
recently demonstrated that the account of the SSE in 2DEG with
equilibrium spin polarization leads to the second branch: the
SEAW. It was done in the longitudinal (the electrostatic limit)
regime. Change of the dispersion dependence of the SEAWs at the
account of the transverse electric field contribution in the wave
propagation which forms the extraordinary SEAW has been
demonstrated.

The influence of the Coulomb exchange interaction on the
propagation of waves in 2DEG has been studied in both regimes: the
single fluid model of electrons and the SSE-QHD.

\section{Acknowledgements}
Work of P.A. is supported by the Russian Foundation for Basic
Research (grant no. 16-32-00886) and by the Dynasty foundation.

\section{\label{sec:level1} Appendix: Integrals for potentials of electromagnetic field}

\begin{equation}\label{2DExWsse } \delta \varphi=q_{e}\sum_{s=u,d}\int\frac{\delta n_{s}(\textbf{r}',t-|\textbf{r}-\textbf{r}'|/c)}{|\textbf{r}-\textbf{r}'|}d\textbf{r}' \end{equation}
\begin{equation}\label{2DExWsse } \delta \textbf{A}=\frac{q_{e}}{c}\sum_{s=u,d}n_{0s}\int\frac{\delta \textbf{v}_{s}(\textbf{r}',t-|\textbf{r}-\textbf{r}'|/c)}{|\textbf{r}-\textbf{r}'|}d\textbf{r}' \end{equation}

We consider the plane wave perturbations $\delta f=F
e^{-\imath\omega t+\imath \textbf{k} \textbf{r}}$ which leads to
$$\int\frac{\delta f(\textbf{r}',t-|\textbf{r}-\textbf{r}'|/c)}{|\textbf{r}-\textbf{r}'|}d\textbf{r}'=$$
$$=F\int\frac{e^{(-\imath\omega(t-|\textbf{r}-\textbf{r}'|/c)+\imath \textbf{k}\textbf{r}')}}{|\textbf{r}-\textbf{r}'|}d\textbf{r}'=$$
$$=F e^{-\imath\omega t+\imath \textbf{k}\textbf{r}} \int\frac{e^{(\imath\omega(|\textbf{r}-\textbf{r}'|/c)+\imath \textbf{k}(\textbf{r}'-\textbf{r}))}}{|\textbf{r}-\textbf{r}'|}d\textbf{r}'=$$
\begin{equation}\label{2DExWsse }=\delta f \int e^{\imath\omega\xi/c+\imath k\xi\cos\varphi} d\varphi d\xi=\Im\delta f ,\end{equation}
where
\begin{equation}\label{2DExWsse Im def}\Im\equiv\int e^{\imath\omega\xi/c+\imath k\xi\cos\varphi} d\varphi d\xi.\end{equation}
As the result of calculation of integral (\ref{2DExWsse Im def})
we find
\begin{equation}\label{2DExWsse }\Im=\frac{2\pi}{k}\frac{1}{\sqrt{1-\frac{\omega^{2}}{k^{2}c^{2}}}}.\end{equation}
Substituting our results in the expressions for $\delta\varphi$
and $\delta \textbf{A}$ we have
\begin{equation}\label{2DExWsse } \delta\varphi=q_{e}\Im (\delta n_{u}+\delta n_{d}),\end{equation}
\begin{equation}\label{2DExWsse } \delta \textbf{A}=\frac{q_{e}}{c}\Im (n_{0u}\delta \textbf{v}_{u}+n_{0d}\delta \textbf{v}_{d}).\end{equation}

\end{document}